\documentclass[12pt]{article}
\usepackage{amssymb}
\usepackage{graphicx}
\def\e{\varepsilon}
\begin{document}

\begin{titlepage}
\begin{center}
{\huge Models for optical solitons in the two-cycle regime}
\end{center}
\vspace{1.5cm}

\noindent

H. Leblond and F. Sanchez
\\\vspace{1cm}

 {\it \noindent
Laboratoire POMA, UMR 6136, Universit\'e d'Angers, 2 Bd Lavoisier,
49000 Angers, France}

\vspace{3cm}

{\bf Abstract}\vspace{5mm}

 We derive model equations for optical pulse propagation in a medium described
 by a two-level Hamiltonian, without the use of the slowly varying envelope approximation.
 Assuming that the resonance frequency of the two-level atoms is either well above or well
 below the inverse of the characteristic duration of the pulse, we reduce the
 propagation problem to a modified Korteweg-de Vries or a sine-Gordon equation.
 We exhibit analytical solutions of these equations which are rather close in shape and spectrum to
 pulses in the two-cycle regime  produced experimentally, which shows that soliton-type propagation of the
 latter can be envisaged.
\end{titlepage}

\section{Introduction}
Recent advances in dispersion managing now allow  the generation of ultrashort optical pulses
containing few oscillations directly from a laser source. Two-cycle pulses have been recently
reported in mode-locked Ti-sapphire lasers using double-chirped mirrors~\cite{fs1,fs2,fs3}. Because the
pulse duration becomes close to the optical period, a question of interest is to know if few
cycle pulses can generate optical solitons in nonlinear media. The usual description of short
pulses propagation in nonlinear optics is made using the nonlinear Schr\"odinger (NLS)
equation which is derived using the slowly varying envelope approximation. However, for
ultrashort pulses considered in this paper the slowly varying envelope approximation is not
valid any more.
This situation, and the corresponding one in the spatial domain, called `non-paraxial',
have already given rise to several studies.
An approach consists in adding corrective terms to the NLS model~\cite{blac00a}.
This high-order perturbation approach still involves the
 slowly varying envelope
 approximation, and requires cumbersome and difficult computations.
The approach of~\cite{zharov}  allows one to determine the ray trajectories in a very rigorous way,
without any use of the paraxial approximation. However, it still makes use of the slowly varying envelope
 approximation in the time
 domain, and therefore  can hardly be generalized to the problem under consideration in this paper.
It is
preferable to leave completely the concept of envelope. Indeed, it is not adapted when a pulse
is composed of few optical cycles. The aim of this paper is to demonstrate that other
approximations can be envisaged and can also lead to completely integrable equations, and
 support solitons. The basic principle of our work is that a soliton can propagate only when the
 absorption is weak, therefore its characteristic frequency must be far away from the
 absorption range of the material. If it is far below, a long-wave approximation can be
 performed. On the other hand, if it is far above, it will be a short-wave approximation. Both
 approaches are used in this paper leading to completely integrable systems. It is organized as
 follows. In section 2 we develop the model which is based on a non absorbing homogeneous
 and isotropic two-level medium. A semi-classical approach is used leading to the well-known
 Maxwell-Bloch equations. The long-wave approximation is investigated in section 3. In this
 case the model reduces to a modified Korteweg-de Vries (mKdV) equation. The two-soliton solution is
 very close to the experimentally observed two-cycle pulses. In section 4 we investigate the
 short-wave approximation. The resulting model is formally equivalent to that describing
 the self-induced transparency. It can be reduced to the sine-Gordon equation. Again, the two-soliton
  solution is comparable with the experimental observations~\cite{fs3}.

\section{Model}

In this section we derive the starting equations for further analysis.
 The medium is treated using the density matrix formalism and the field using the Maxwell
equations.

We consider an homogeneous medium, in which the dynamics of each
atom is described by a two-level Hamiltonian
\begin{equation}
H_0=\hbar\left(
\begin{array}{cc}
\omega_a & 0 \\
0 & \omega_b
\end{array}
\right).
\end{equation}
A more realistic description should take into account an arbitrary number of atomic levels.
Indeed, we consider wave frequencies far  from the resonance line of the medium, and in this situation all
transitions should be taken into account. But we intend here to suggest a new approach to the
 description of ultrashort optical pulses. Therefore we restrict the study to a very simple
 and rather academic model.

The atomic dipolar electric momentum is assumed to be along the $x$-axis. It is thus described by the operator
$\vec\mu=\mu\vec e_x$, where $\vec e_x$ is the unitary vector along the $x$-axis
and
\begin{equation}
\mu=\left(
\begin{array}{cc}
0 & \mu \\
\mu^\ast & 0
\end{array}
\right).\label{deux}
\end{equation}
The polarization
density $\vec P$ is related to the
density matrix $\rho$ through
\begin{equation}
\vec P=N \mbox{tr}(\rho\vec \mu), \label{3}
\end{equation}
where $N$ is the number of atoms per unit volume. Thus $\vec P$ reduces to $P\vec e_x$.

 The
electric field $\vec E$ is governed by the Maxwell equations.
In the absence of magnetic effects, and assuming that the wave is a plane wave propagating along the $z$-axis,
polarized along the $x$-axis, $\vec E=E\vec e_x$, they reduce to
\begin{equation}
\partial_z^2 E=\frac{1}{c^2}\partial_t^2( E+4\pi P).  \label{M}
\end{equation}
 $c$ is the light velocity in vacuum. We denote by
$\partial_t$ the derivative operator $\frac\partial{\partial t}$
with regard to the time variable $t$, and  so on.

The coupling between the
atoms and the electric field is taken into account by a coupling
energy term in the total Hamiltonian $H$, that reads:
\begin{equation}
H= H_0-\mu E.  \label{4}
\end{equation}
The density matrix evolution equation (Schr\"odinger equation) writes as
\begin{equation}
i\hbar\partial_t\rho=\left[H,\rho\right]+{\cal R},  \label{S}
\end{equation}
where $\cal R$ is some phenomenological relaxation term.
The set of equations (\ref{M}-\ref{S}) is sometimes called the
Maxwell-Bloch equations, although this name denotes more often a
reduction of it.

Setting
\begin{equation}
 t'=ct\quad,\quad P'=4\pi P\quad,\quad\rho'=4\pi N\hbar c\rho\quad,\quad\mu'=\frac\mu{\hbar c},
 \label{norm}
\end{equation}
\begin{equation}
H'=\frac H{\hbar c}
\quad,\quad
H_0'=\frac{H_0}{\hbar c}
\quad,\quad
\omega_{a,b}'=\frac{\omega_{a,b}}{ c},
\end{equation}
allows one to replace the constants $c$, $N$, $\hbar$ and $4\pi$ in  system (\ref{3}-\ref{S}) by 1.
We denote the components of $\rho$ by
\begin{equation}
\rho=\left(\begin{array}{cc}\rho_{a}&\rho_{t}\\\rho_{t}^*&\rho_{b}\end{array}\right),
\end{equation}
and so on,
and by $\Omega=\omega_b-\omega_a$  the resonance frequency of the atom.

The relaxation expresses as
\begin{equation}
{\cal R}=i\hbar \left(\begin{array}{cc}
\rho_b/\tau_b&-\rho_t/\tau_t\\
-\rho_t^*/\tau_t&-\rho_b/\tau_b\end{array}\right),
\end{equation}
where $\tau_b$ and $\tau_t$ are the relaxation times for the populations and for the coherences respectively.
We show below that, according to the fact that relaxation occurs very slowly with regard to optical oscillations,
the relaxation term $\cal R$ could be omitted.

\section{Long-wave approximation}
\subsection{A modified Korteweg-de Vries equation}
Let us  first consider the situation where the wave duration $t_w$ is long with regard to
 the period $t_r=2\pi/\Omega$ (recall that $\Omega=\omega_b-\omega_a$)
  that corresponds to the resonance frequency of the two-level
atoms. We assume that $t_w$ is about one optical period, say about
one femtosecond. Thus we assume that the resonance frequency
$\Omega$ is large with regard to optical frequencies. In order to
obtain soliton-type propagation, nonlinearity must balance
dispersion, thus the two effects must arise simultaneously in the
propagation. This involves a small amplitude approximation.
Further, we can speak of soliton only if  the pulse shape is kept
 on a large propagation distance.
Therefore we use the reductive perturbation method as defined in \cite{tanw1}.
We expand the electric field $E$, the polarization density $P$ and  the density matrix
 $\rho$ as power series of a small
parameter $\e$ as
\begin{equation}
E=\sum_{n\geqslant 1}\e^nE_n\quad,\quad
P=\sum_{n\geqslant 1}\e^nP_n\quad,\quad\rho=\sum_{n\geqslant 0}\e^n\rho_n,\label{lw1}
\end{equation}
and introduce the slow variables
\begin{equation}
\tau=\e\left(t-\frac zV\right)\quad,\quad
\zeta=\e^3 z.
\end{equation}
Expansion (\ref{lw1}) gives an account of the small amplitude
approximation. The retarded time variable $\tau$  describes the pulse shape, propagating at speed $V$ in a
first approximation. Its order of magnitude $\e$ gives account for
the long-wave approximation, so that the pulse duration $t_w$ has the same order of magnitude
as $t_r/\e$.
The propagation distance is assumed to be very long with regard to the pulse length $ct_w$,
therefore it will have the same order of magnitude as $ct_r/\e^n$, where $n\geqslant 2$.
The  value of $n$ is determined by the distance at which dispersion effects occur. According to the general theory
of the derivation of KdV-type equations~\cite{tanw1}, it is $n=3$.
The $\zeta$ variable of order $\e^3$ describes thus long-distance propagation.
The physical values of the relaxation times  $\tau_b$ and $\tau_t$ are in the picosecond range, or even slower,
thus very large with regard to the pulse duration $t_w$.
Therefore we write
\begin{equation}
\tau_j=\frac{\hat\tau_j}{\e^2}\quad\mbox{ for $j=b$ and~$t$.}
\label{relax1}
\end{equation}

The Schr\"odinger equation (\ref{S}) at order $\e^0$ is satisfied
by the following value of
$\rho_0$, which represents a steady state in which all atoms are in their fundamental state $a$:
\begin{equation}
\rho_0=\left(\begin{array}{cc}\alpha&0\\0&0\end{array}\right).
\end{equation}
Notice that, according to the change of variables  (\ref{norm}), the trace $\mbox{tr}(\rho)$ of the density matrix
is not 1 but $\alpha=4\pi N\hbar c$.
Then the Schr\"odinger equation (\ref{S}) at order $\e^1$
yields
\begin{equation}
\rho_{1t}=\frac{\mu\alpha}\Omega E_1,
\end{equation}
so that
\begin{equation}
P_1=\frac{2\vert\mu\vert^2\alpha}{\Omega}E_1.
\end{equation}
The Maxwell equation (\ref{M}) at order $\e^3$ gives the value of the velocity
\begin{equation}
V=\left(1+\frac{2\vert\mu\vert^2\alpha}\Omega\right)^{\frac{-1}2},
\end{equation}
in accordance with the limit of the dispersion relation as the frequency $\omega$ tends to zero.

The Schr\"odinger equation (\ref{S}) at order $\e^2$ yields
$\rho_{1a}=\rho_{1b}=0$
and
\begin{equation}
\rho_{2t}=\frac{\mu\alpha}\Omega E_2-\frac{i\mu\alpha}{\Omega^2}\partial_\tau E_1.
\end{equation}
Then
\begin{equation}
P_2=\frac{2\vert\mu\vert^2\alpha}{\Omega}E_2,
\end{equation}
and the Maxwell equation (\ref{M}) at order $\e^4$ is automatically satisfied.

The Schr\"odinger equation (\ref{S}) at order $\e^3$ gives
\begin{equation}
\rho_{2b}=-\rho_{2a}=\frac{\vert\mu\vert^2\alpha}{\Omega^2}E_1^2
\end{equation}
and
\begin{equation}
\rho_{3t}=\frac{\mu\alpha}\Omega E_3-\frac{i\mu\alpha}{\Omega^2}\partial_\tau E_2
-\frac{\mu\alpha}{\Omega^3}\partial_\tau^2 E_1-\frac{2\mu\vert\mu\vert^2\alpha}{\Omega^3} E_1^3
-\frac{i\mu\alpha}{\Omega^2\tau_t}E_1.
\end{equation}

The corresponding term of the  polarization density $P$
contains a nonlinear term:
\begin{equation}
P_3=\frac{2\vert\mu\vert^2\alpha}\Omega E_3-\frac{2\vert\mu\vert^2\alpha}{\Omega^2}\partial_\tau^2 E_1
-\frac{4\vert\mu\vert^4\alpha}{\Omega^3}E_1^3,
\end{equation}
but the terms involving the relaxation do not appear.
The Maxwell equation at order $\e^5$ yields the following evolution equation for the main electric field amplitude $E_1$:
\begin{equation}
\partial_\zeta E_1=\frac{V\vert\mu\vert^2\alpha}{\Omega^3}\partial_\tau^3E_1+
\frac{2V\vert\mu\vert^4\alpha}{\Omega^3}\partial_\tau E_1^3,\label{mkdv}
\end{equation}
which is a mKdV equation.
Equation (\ref{mkdv}) can be generalized as follows:
a general derivation of KdV-type models \cite{linkdv} shows that the  coefficient
of the dispersive term $\partial_\tau^3E_1$ in this equation must be $(1/6){d^3k}/{d\omega^3}$.
We check by direct computation of the dispersion relation that it holds in the present case.
Another heuristic reasoning can relate the value of the nonlinear coefficient of equation
(\ref{mkdv}) to the third order nonlinear susceptibility $\chi^{(3)}$. It uses the nonlinear Schr\"odinger (NLS)
equation which describes the evolution of a short pulse envelope  in the same medium.
The NLS equation writes as (\cite{boyd}, 6.5.32)
\begin{equation}
i\partial_\zeta {\cal E}-\frac12k_2\partial_\tau^2{\cal E}+\gamma{\cal E}\left\vert{\cal E}\right\vert^2=0,
\end{equation}
where $\cal E$ is the envelope amplitude of the wave electric field,
 $k_2$ the group velocity dispersion, and $\gamma$ is related to
$\chi^{(3)}$ through
\begin{equation}
\gamma=\frac{6\omega\pi}{nc}\chi^{(3)}.
\end{equation}
$n$ is the refractive index of the medium, $\omega$ the wave pulsation.
We drop the dispersion term, replace $\omega$ by $i\partial_\tau$, and notice that $\cal E$ coincides
with the real field $E_1$  at the long-wave limit, to get
\begin{equation}
\partial_\zeta E_1=\frac{-6\pi}{nc}\chi^{(3)}\partial_\tau E_1^3.\label{nlslim}
\end{equation}
Equation (\ref{nlslim}) gives an expression of the nonlinear coefficient in the mKdV
 equation (\ref{mkdv}).
The relevant component of the third order nonlinear susceptibility tensor $\chi^{(3)}$
computed from the above model
is \cite{boyd}
\begin{equation}
\chi ^{(3)}_{xxxx}(\omega ,\omega ,\omega ,-\omega )=-\frac{4%
}{3}\frac{N}{\hbar ^{3}}\frac{\Omega |\mu |^{4}}{\left( \omega ^{2}-\Omega
^{2}\right) ^{2}}\;.  \label{a=0}
\end{equation}
Taking the long-wave limit $\omega\longrightarrow 0$ in equation (\ref{a=0}), we check that the
expression of the nonlinear coefficient obtained from (\ref{nlslim}) holds in the present case.
Thus we can write equation (\ref{mkdv}) as
\begin{equation}
\partial_\zeta E_1=\frac16\left.\frac{d^3k}{d\omega^3}\right\vert_{\omega=0}\partial_\tau^3E_1
-\left.\frac{6\pi}{nc}\chi^{(3)}_{xxxx}(\omega ,\omega ,\omega ,-\omega )\right\vert_{\omega=0}\partial_\tau E_1^3\;.
\label{genmkdv}
\end{equation}
It can be reasonably conjectured that equation (\ref{genmkdv}) will still hold in the more general case
of an arbitrary number of atomic levels, when the inverse of the characteristic pulse duration is
much smaller than any
of the transition  frequencies of the atoms.
\subsection{The two-soliton solution}
The mKdV equation (\ref{mkdv}) is completely integrable by means
of the inverse scattering transform \cite{wad73}. The $N$-soliton solution has been given by Hirota \cite{hirb76a}.
In order to write it easily, we write the mKdV equation (\ref{mkdv}) into the dimensionless form
\begin{equation}
\partial_Z u+2\partial_T u^3+\partial_T^3u=0,\label{mkdvd}
\end{equation}
where $u$ is a dimensionless electric field, and $Z$ and $T$ dimensionless space and time variables defined by
\begin{equation}
u=\frac{E_1}{E_0}\quad,\quad Z=\frac{-\zeta}L\quad\quad T=\frac\tau{T_0}\label{red1}.
\end{equation}
The characteristic electric field, space and time
are defined by
\begin{equation}
E_0=\frac\Omega{\vert\mu\vert}\quad,\quad
T_0=\frac1\Omega\quad,\quad L=\frac1{\alpha\vert\mu\vert^2V},\label{red2}
\end{equation}
in normalized units.
Relations (\ref{red1}-\ref{red2}) can be expressed in a more convenient way as follows.
 Let us first choose as reference
 time the pulse length $t_w$ (in physical unit).
 The small perturbative parameter is then
 \begin{equation}\e=\frac1{t_w\Omega}.
 \end{equation}
 The characteristic electric field $\cal E$, and propagation distance $\cal L$,
 are
 \begin{equation}
{\cal E}=\frac\hbar{t_w\vert\mu\vert}\quad,\quad
{\cal L}=\frac{\hbar c^2\Omega^3t_w^3}{4\pi NV\vert\mu\vert^2},
\end{equation}
where the speed $V$ is
\begin{equation}\label{vnn}
  V=c\left(1+\frac{8\pi\vert\mu\vert^2N}{\hbar\Omega}\right)^{\frac{-1}2}.
\end{equation}
Then the quantities involved by the dimensionless equation (\ref{mkdvd})
are related to the quantities measured in the laboratory through
 \begin{equation}
u=\frac{E}{\cal E}\quad,\quad Z=\frac{-z}{\cal L}\quad,\quad
 T=\frac 1{t_w}\left(t-\frac zV\right).
\end{equation}

The soliton solution writes as
\begin{equation}
u=p\,\mbox{sech}\,\eta,
\end{equation}
with
\begin{equation}
\eta=p T-p^3Z-\gamma,
\end{equation}
$p$ and $\gamma$ being arbitrary parameters.

\begin{figure}[hbt!]
\begin{center}
\includegraphics[width=9cm]{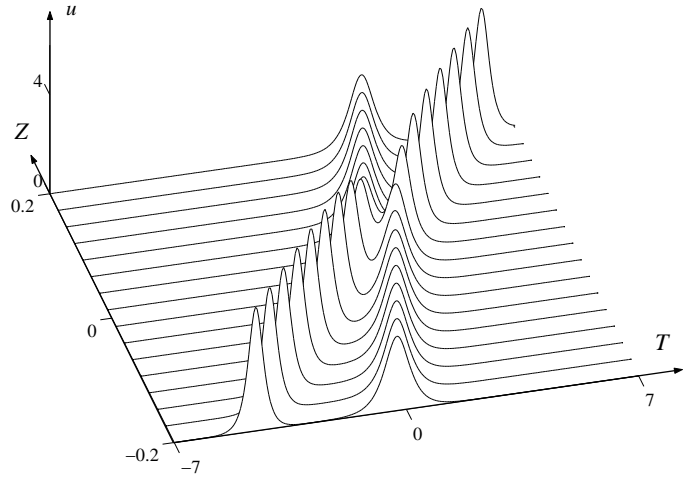}
\caption{\footnotesize   Two-soliton solution of the mKdV equation, using dimensionless parameters.} \label{figmkdv}
\end{center}
\end{figure}

\begin{figure}[hbt!]
\begin{center}
\includegraphics[width=9cm]{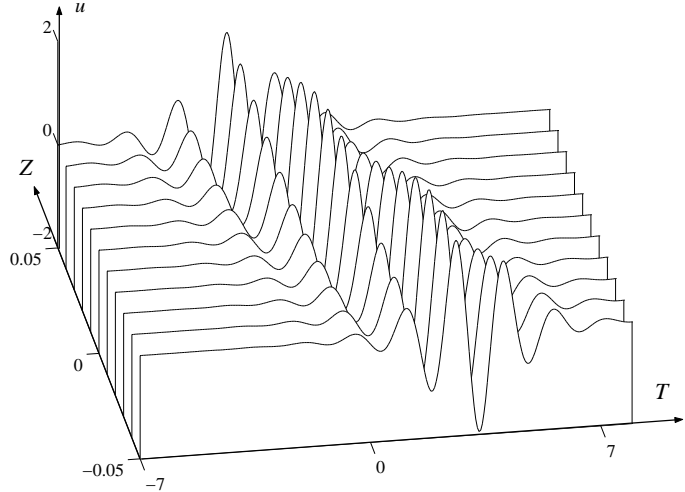}
\caption{\footnotesize   Second-order soliton solution of the mKdV equation,
 using dimensionless parameters.} \label{figmkdv2}
\end{center}
\end{figure}

\begin{figure}[hbt!]
\begin{center}
\includegraphics[width=9cm]{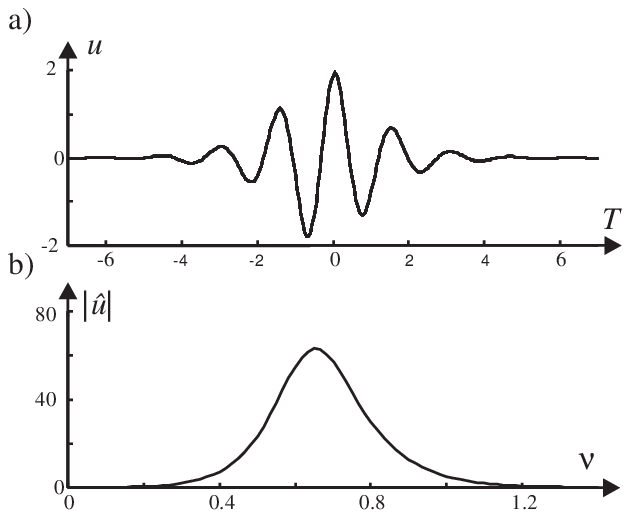}
\caption{\footnotesize
  (a) Pulse profile and (b) spectrum, of the second-order soliton solution of the mKdV equation
   of figure \ref{figmkdv2}. Dimensionless parameters.} \label{figmkdv3}
\end{center}
\end{figure}

The two-soliton solution is
\begin{equation}
u=\frac{e^{\eta_1}+e^{\eta_2}+
\left(\frac{p_1-p_2}{p_1+p_2}\right)^2
\left(\frac{e^{\eta_1}}{4p_1^2}+\frac{e^{\eta_2}}{4p_2^2}\right)e^{\eta_1+\eta_2}}
{1+\frac{e^{2\eta_1}}{4p_1^2}+\frac2{(p_1+p_2)^2}e^{\eta_1+\eta_2}+\frac{e^{2\eta_2}}{4p_2^2}
+\left(\frac{p_1-p_2}{p_1+p_2}\right)^4\frac{e^{2\eta_1+2\eta_2}}{16p_1^2p_2^2}}\;,\label{mkdvsol}
\end{equation}
with
\begin{equation}
\eta_j=p_j T-p_j^3Z-\gamma_j\,,
\end{equation}
for $j=1$ and 2. The parameters $p_1$, $p_2$, $\gamma_1$ and $\gamma_2$ are arbitrary.
When they take real values, the explicit solution
(\ref{mkdvsol}) describes the interaction of two localized bell-shaped pulses, which are solitons.
An example of this solution is drawn on figure \ref{figmkdv}, using the values of the  parameters $p_1=3$,
$p_2=5$, $\gamma_1=\gamma_2=0$.
But expression (\ref{mkdvsol}) also describes the so-called higher order solitons, which can
be considered as a pair of solitons of the above kind linked together, and have often an oscillatory behaviour.
An example is given on figure \ref{figmkdv2}. It uses the values of parameters
$p_1=1+4i$, $p_2=p_1^\ast$, $\gamma_1=-\gamma_2=i\pi/2$. The corresponding
spectrum is drawn on figure \ref{figmkdv3}.

These spectrum and pulse profile  are comparable to the experimental pulses given by \cite{fs3}.
It can thus be thought that the two-cycle pulses produced experimentally could
propagate as solitons in certain media, according to the mKdV model.

\section{Short-wave approximation}
\subsection{A sine-Gordon equation}

We now consider the situation in which the resonance frequency $\Omega$ of the atoms is below the optical frequencies.
Then the characteristic pulse duration $t_w$ is very small with regard to $t_r=2\pi/\Omega$,
thus we use a short-wave approximation.
We introduce a small perturbative parameter $\e$,
such that the resonance period $t_r=\hat t_r/\e$, where $\hat t_r$ has the same order of magnitude
as  the pulse duration $t_w$.
The perturbative parameter $\e$ is thus about $t_w/t_r$.
Consequently, the Hamiltonian  $H_0$ of the atom
is replaced in the Schr\"odinger
equation (\ref{S}) by
\begin{equation}
\e \hat H_0.\label{swh1}
\end{equation}
We introduce a retarded time $\tau$ and a slow propagation variable $\zeta$
such that
\begin{equation}
\tau=\left(t-\frac zV\right)\quad,\quad \zeta=\e z.\label{swvar1}
\end{equation}
The zero order reference time is chosen to be $t_w$, therefore $\tau$ is not a slow variable.
The definition of the variable $\zeta$ gives account for long distance propagation.
Computation shows that the dispersion effects arise at distances about $ct_w/\e$, from
which follows the choice of the order of magnitude  of $\zeta$.
The electric field $E$ is expanded as
$E=\sum_{n\geqslant0}\e^nE_n$, and so on.
The  pulse duration $t_w$ is still assumed to be about one femtosecond, corresponding to  an optical pulse of a few cycles.
The relaxation times $\tau_b$, $\tau_t$ are very long with regard to $t_w$. Since the above scaling uses
$t_w$ as zero-order reference time, this can be expressed by setting
\begin{equation}
\tau_j=\frac{\hat\tau_j}{\e}\quad\mbox{for $j=b$ and~$t$.}
\label{relax2}
\end{equation}
Notice that (\ref{relax2}) differs formally from the assumption (\ref{relax1}) written in the previous section
but represents the same physical hypothesis.

The above scaling can also  be presented from another viewpoint, taking
the characteristic time of the resonance $1/\Omega$ as zero-order reference time,  as follows.
The relevant component of the third order nonlinear susceptibility tensor $\chi^{(3)}$
computed from the above model
is given by formula (\ref{a=0}),
where $\omega$ is the wave frequency (while $\Omega=\omega_b-\omega_a$ is the resonance frequency of the two-level system).
The short-wave approximation corresponds to $\omega\longrightarrow \infty$. Then
$\chi ^{(3)}_{xxxx}$ tends to zero. Thus a linear behaviour of the wave can be expected in the
short-wave approximation, except if the nonlinearity is very strong. The latter physical assumption
can formally be expressed by assuming that  the product $\mu E$ is very large with regard to $\Omega$,
according to
\begin{equation}
H=H_0+\frac1 \e\mu E,\label{swh2}
\end{equation}
where the small parameter $\e$ tends to zero.
Then the short wave approximation can be sought using the expansions
\begin{equation}\rho=\sum_{j\geqslant0}\e^j\rho_j\quad,\quad
E=\sum_{j\geqslant0}\e^jE_j\quad,\quad
P=\sum_{j\geqslant0}\e^j P_j,
\end{equation}
and slow variables $\zeta$ and $\tau$ such that
\begin{equation}
\partial_t=\frac1\e\partial_\tau\quad,\quad\partial_z=\frac{-1}{\e V}\partial_\tau+\partial_\zeta.\label{varsw2}
\end{equation}
The definition of variables (\ref{varsw2}) is very close to the standard short wave approximation formalism
developed {\it e.g.} in \cite{kra00a,man01a}. It is easily checked that the scalings (\ref{swh1}-\ref{swvar1})
and (\ref{swh2}-\ref{varsw2}) are equivalent. We refer to the former below.

The Schr\"odinger equation (\ref{S}) at order $\e^0$ yields
\begin{eqnarray}
i\partial_\tau\rho_{0a}=-E_0\left(\mu \rho_{0t}^\ast-\mu^\ast\rho_{0t}\right),\label{S-1a}\\
i\partial_\tau\rho_{0b}=+E_0\left(\mu \rho_{0t}^\ast-\mu^\ast\rho_{0t}\right),\label{S-1b}\\
i\partial_\tau\rho_{0t}=-E_0\mu\left(\rho_{0b}-\rho_{0a}\right).\label{S-1t}
\end{eqnarray}
From (\ref{S-1a}-\ref{S-1b}) we retrieve the normalization condition of the density matrix
$\partial_\tau \mbox{tr}\, \rho=0$.
We introduce the population inversion $w=\rho_{0b}-\rho_{0a}$
and get
\begin{equation}
\rho_0=\left(\begin{array}{cc}(\alpha-w)/2&i\mu\int^\tau E_0w\vspace{1.5mm}\\
-i\mu^\ast\int^\tau E_0w&(\alpha+w)/2\end{array}\right)
\end{equation}
(as above, $\mbox{tr} \rho=\alpha=4\pi N\hbar c$ due to the normalization), and the equation
\begin{equation}
 \partial_\tau w=-4\vert\mu\vert^2 E_0\int^\tau E_0w.\label{sw1}
\end{equation}

Then expression (\ref{3}) of the polarization $P$ yields $P_0=0$,
and the Maxwell equation (\ref{M}) at order $\e^0$ becomes trivial if the velocity is chosen as $V=1$.

The Schr\"odinger equation (\ref{S}) at order $\e$
writes then as
\begin{equation}
i\partial_\tau\rho_1=\left[H_0,\rho_0\right]
-\left[\mu E_0,\rho_1\right]-\left[\mu E_1,\rho_0\right]+
i\left(\begin{array}{cc}\rho_{0b}/{\tau_b}.&-\rho_{0t}/{\tau_t}\\
-\rho_{0t}^*/{\tau_t}&-\rho_{0b}/{\tau_b}\end{array}\right).\label{S0}
\end{equation}
Defining $w_1=\rho_{1b}-\rho_{1a}$, the off-diagonal components of equation
(\ref{S0}) yield
\begin{equation}
\rho_{1t}=i\left(\Omega+\frac i{\tau_t}\right)\int^\tau\rho_{0t}+i\mu\int^\tau(E_0w_1+E_1w),
\end{equation}
so that the corresponding term $P_1=\mu^\ast\rho_{1t}+\mu\rho_{1t}^\ast$
of the polarization is
\begin{equation}
P_1=-2\Omega\left\vert\mu\right\vert^2\int^\tau\int^\tau Ew.
\end{equation}
Notice again that the relaxation does not appear in the expression of the polarization at this order.
The Maxwell equation (\ref{M}) at order $\e$ then reduces to
\begin{equation}
\partial_\zeta\partial_\tau E_0=\Omega\vert \mu\vert^2 E_0w \;.\label{sw2}
\end{equation}
Equations (\ref{sw1},\ref{sw2}) yield the sought system.
If we set
\begin{equation}
p=-i\vert\mu\vert^2\int^\tau E_0w,
\end{equation}
they reduce to
\begin{eqnarray}
\partial_\zeta E_0&=&i\Omega p,\label{sw3}\\
\partial_\tau p&=&-i\vert\mu\vert^2 E_0w,\label{sw4}\\
\partial_\tau w&=&-4iE_0p,\label{sw5}
\end{eqnarray}
which coincide with the equations of the self-induced transparency,
although the physical situation is quite different:
the characteristic frequency $1/t_w$ of the pulse is far above the resonance frequency  $\Omega$,
while the self-induced transparency occurs when the optical field oscillates at the frequency
$\Omega$. The quantities $E$ and $w$ describe here
 the electric field and population inversion themselves, and not amplitudes modulating a carrier
 with frequency  $\Omega$. Notice that $E$ and $w$ are here real quantities, and not complex ones as
 in the case of the self-induced transparency. Further,
 $p$ is not the polatization density, but is proportional to its $\tau$-derivative.
  Another difference is the absence of a factor $1/2$ in the right-hand side of equation (\ref{sw3}).

  Since they explicitly involve the population inversion, the model equations (\ref{sw3}-\ref{sw5})
cannot be generalized easily to more realistic situation in which an arbitrary number of atomic levels
are taken into account. Recall that, according to the assumption made at the beginning of the section,
this model is valid  for a very strong nonlinearity only.
In particular, we assumed that the atomic dipolar momentum $\mu$ has a very large value.
In a more realistic situation, it can be expected that only the transition corresponding to the largest
value of the dipolar momentum will have a significant contribution.
If several transitions correspond to large values of the dipolar momentum with the same order of magnitude,
we can expect that
the short-wave approximation will yield some more complicated asymptotic system involving
the populations of each level concerned. The derivation of such a model is left for further study.

  \subsection{The two-soliton solution}
  Using dimensionless variables defined by
  \begin{equation}
  \Theta=\frac{E_0}{E_r}\quad,\quad W=\frac w{w_r}
  \quad,\quad T=\frac\tau{T_0}\quad,\quad Z=\frac\zeta L,\label{swsol1}
\end{equation}
where the reference values
satisfy
\begin{equation}
E_rT_0\vert\mu\vert=1\quad\mbox{and}\quad
w_rL\Omega T_0\vert\mu\vert^2=2,
\end{equation}
and setting $\eta=\int^ZW$,
the system (\ref{sw1}-\ref{sw2}) reduces to
\begin{eqnarray}
\partial_Z\partial_T\Theta&=&2\Theta\partial_Z\eta,\label{swkra1}\\
\partial_Z\partial_T\eta&=&-2\Theta\partial_Z\Theta\label{swkra2}.
\end{eqnarray}
Equations (\ref{swkra1}-\ref{swkra2}) have been found to describe short electromagnetic wave propagation in ferrites,
using the same kind of short-wave approximation~\cite{kra00a}.

The following change of dependant variables:
\begin{eqnarray}
\partial_Z\eta&=&A\cos u\label{swch1},\\
\partial_Z\Theta&=&A\sin u\label{swch2},
\end{eqnarray}
 transforms equations (\ref{swkra1}-\ref{swkra2}) into \cite{kra00a,ablo2}
\begin{eqnarray}
\partial_TA&=&0,\label{sG1}\\
\partial_Z\partial_T u&=&2A\sin u\label{sG2}.
\end{eqnarray}
Since, according to (\ref{sG1}), $A$ is a constant, (\ref{sG2}) is the sine-Gordon equation.
Before we recall some properties of the latter, let us determine  the physical meaning of the  constant $A$ in
the present physical frame.
Using relations (\ref{swch1}-\ref{swch2}) and the definition of $\eta$, we find that
\begin{equation}
A^2=\lim_{T\longrightarrow \infty}\left(W^2+\left(\partial_Z\Theta\right)^2\right)
\end{equation}
Since $\Theta$ is the dimensionless wave electric field, it vanishes at infinity.
Thus we can have a non-vanishing solution only if some initial population inversion $w_i=W_iw_r$
is present. The constant involved by equation (\ref{sG2}) is then $A=W_i$.

Using the variable ${\hat Z}=2W_iZ$, equation (\ref{sG2}) reduces to the sine-Gordon equation
\begin{equation}
\partial_{\hat Z}\partial_T u=\sin u\label{sG3}.
\end{equation}
The quantities involved by equation (\ref{sG3}) are related to the quantities measured in the laboratory
through
\begin{eqnarray}
\hat Z=\frac z{\hat L} \quad,\quad T=\frac1{t_w}\left(t-\frac zc\right),\\
E=\frac{E_r}2\int^{\hat Z}\sin u\quad,\quad
w=w_i \cos u.
\end{eqnarray}
The electric field and propagation length scaling parameters are
\begin{eqnarray}
E_r&=&\frac\hbar{\vert\mu\vert t_w},\\
\hat L&=&\frac{\hbar c}{\Omega t_w4\pi N\vert\mu\vert^2w_i},\label{swlength}
\end{eqnarray}
in which the initial population inversion $w_i$ and typical pulse duration $t_w$ are given.
The small perturbative parameter $\e$ can be identified with $ \Omega t_w$, expressing the fact that
$t_w$ is very small with regard to $1/\Omega$.

\begin{figure}[hbt!]
\begin{center}
\includegraphics[width=9cm]{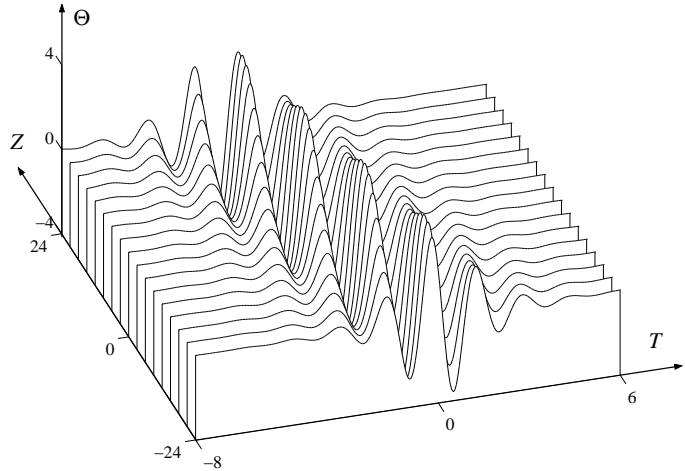}
\caption{\footnotesize   Electric field evolution of the second-order soliton solution of the sine-Gordon equation,
 using dimensionless parameters.} \label{figsinG1}
\end{center}
\end{figure}

\begin{figure}[hbt!]
\begin{center}
\includegraphics[width=9cm]{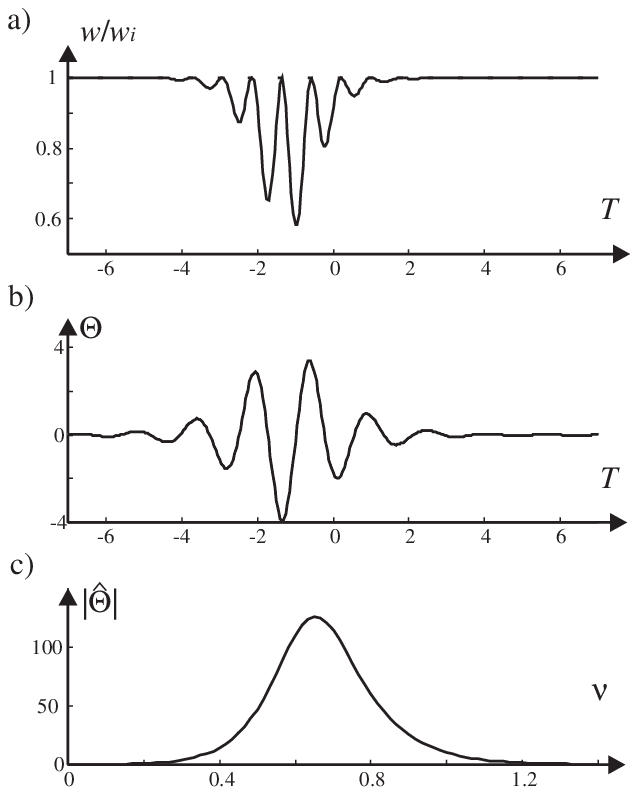}
\caption{\footnotesize
 (a) Pulse profile, (b) population inversion  and (c) spectrum, of the second-order soliton
 solution of the sine-Gordon equation
   of figure \ref{figsinG1}. Dimensionless parameters.} \label{figsinG2}
\end{center}
\end{figure}

The sine-Gordon equation (\ref{sG3}) is completely integrable \cite{ablo2}. A $N$-soliton solution can be found
using either the IST or the Hirota method. As in section 2, we will consider here the two-soliton solution
only, which is \cite{ablo2}
\begin{equation}
u=2i \ln \left(\frac{f^\ast}f\right)\label{sg2sol1},
\end{equation}
with
\begin{equation}
f=1+ie^{\eta_1}+ie^{\eta_2}-\frac{(k_1-k_2)^2}{(k_1+k_2)^2}e^{\eta_1+\eta_2}\label{sg2sol2},
\end{equation}
where
\begin{equation}
\eta_j=k_jT+\frac Z{k_j}+\gamma_j\quad\mbox{for $j=1$, $2$,}\label{sg2sol3}
\end{equation}
$k_1$, $k_2$, $\gamma_1$ and $\gamma_2$ being arbitrary parameters.
When they take real values, formulas (\ref{sg2sol1}-\ref{sg2sol3}) describe the  interaction
of two solitons. The behaviour is very close to that of the  typical two-soliton solution of the
mKdV equation shown
in figure \ref{figmkdv}.
As in the case of the long wave approximation, the two-soliton solution  (\ref{sg2sol1}-\ref{sg2sol3})
is also able to describe soliton-type propagation of a pulse in the two-cycle regime.
The corresponding analytic solution is a second-order soliton or breather, which can be considered
as two bounded solitons, and is obtained using complex conjugate values of the soliton parameters
$k_1$ and $k_2$.
An  example is given in figure \ref{figsinG1}, with the values of parameters
$k_1=1+4i$, $k_2=1-4i$, $\gamma_1=\gamma_2=0$.
The pulse profile, with the corresponding population inversion and spectrum are drawn in figure \ref{figsinG2}.
The profile and spectrum are comparable with the experimental observation of \cite{fs3}.
Notice again  that an initial population inversion $w_i\neq0$ is required.
Total inversion ($w_i=1$) is not necessary but, as shows the expression  (\ref{swlength})
of the propagation reference length $\hat L$, a small inversion reduces the soliton amplitude
and increases the propagation distance at which nonlinear effects occur.
\section{Conclusion}
We have given two models that allow the description of ultrashort optical pulses propagation
in a medium described by a two-level Hamiltonian,
when the slowly varying envelope approximation cannot be used.
Using approximations based on the hypothesis that the resonance frequency of the medium is far from the
field frequency, we derived completely integrable models.
When the resonance frequency is well above the inverse of the typical pulse width of about one femtosecond,
a long-wave approximation leads to a mKdV equation.
When in the contrary the resonance frequency is well below the field frequency, a short-wave  approximation
leads to a model formally identical to that describing self-induced transparency,
but in very different validity conditions. It can be reduced to the sine-Gordon equation.
The scaling parameters for these approximations have been written down explicitly.

Both the mKdV and the sine-Gordon equations are completely integrable by means of the IST method and
admit $N$-soliton solutions.
The two-soliton solution is able to describe the propagation of a pulse in the two-cycle regime,
very close in shape and spectrum to the pulses of this type produced experimentally.
It does not mean that the formulas of this paper describe the experimental results,
because we have considered a propagation problem, and experimental results concern pulses
generated
directly at the laser output.
But we have shown that soliton-type propagation, with only periodic deformation of the pulse
during the propagation, may occur for such type of pulses, under adequate conditions.
In the short-wave approximation, these conditions involve an initial population inversion, at least a partial one.

Further, the study of a two-level Hamiltonian can be considered as an academic problem,
showing the tractability of such an approach. A rather remarkable feature is that the
computations involved  by the derivation of the asymptotic models are relatively short and easy.
Therefore the application of the same approach to more realistic situations can be reasonably envisaged.
A generalization of the mKdV equation obtained in the long-wave approximation has been proposed on heuristic grounds,
and should be justified rigorously. A  generalization of the model obtained in the short-wave approximation
would require a special study. Last, in a more realistic model, it can be envisaged that some transition frequencies
are well above the inverse of the characteristic pulse duration, but that some other are below it.
The treatment of such a situation will mix the above short-wave and long-wave approximations,
 it is left for further study. It can be expected that the result will depend strongly on the
 particular physical situation considered.

\section*{Acknowledgments}
The authors thank M.A. Manna (Universit\'e de Montpellier, France) and
R.A. Kraenkel (University of S\~ao Paulo, Brazil) for fruitful scientific discussions.

\end{document}